\documentclass[aps,prb,twocolumn,superscriptaddress,showpacs,preprintnumbers]{revtex4-1}
\usepackage{natbib}
\usepackage[english]{babel}
\usepackage[dvips]{graphics}
\usepackage{graphicx,epsfig}
\usepackage{amsmath}
\usepackage{color}
\usepackage{multirow}
\usepackage[normalem]{ulem}
\usepackage{booktabs}
\usepackage{physics}
\usepackage{dsfont}
\usepackage{amssymb}
\usepackage{lineno}
\begin{document}
\title{Taking advantage of noise in quantum reservoir computing}
%
\author{L. Domingo}
\email[E--mail address: ]{laia.domingo@icmat.es}
\affiliation{Instituto de Ciencias Matemáticas (ICMAT); Campus de Cantoblanco; 
Nicolás Cabrera, 13-15; 28049 Madrid, Spain}
\affiliation{Departamento de Química; Universidad Autónoma de Madrid;
Cantoblanco - 28049 Madrid, Spain}
\affiliation{Grupo de Sistemas Complejos; Universidad Politécnica de Madrid; 
28035 Madrid, Spain}

\author{G. Carlo}
\email[E--mail address: ]{carlo@tandar.cnea.gov.ar}
\affiliation{Comisi\'on Nacional de Energ\'ia At\'omica, CONICET, 
Departamento de F\'isica, Av.\ del Libertador 8250, 1429 Buenos Aires, Argentina}
\author{F. Borondo}
\email[\textit{Corresponding author}, e--mail address: ]{f.borondo@uam.es}
\affiliation{Instituto de Ciencias Matemáticas (ICMAT); Campus de Cantoblanco;
Nicolás Cabrera, 13-15; 28049 Madrid, Spain}
\affiliation{Departamento de Química; Universidad Autónoma de Madrid;
Cantoblanco - 28049 Madrid, Spain}

\date{\today}
\begin{abstract}
The biggest challenge that quantum computing and quantum machine learning are currently 
facing is the presence of noise in quantum devices. 
As a result, big efforts have been put into correcting or mitigating the induced errors.
But, can these two fields benefit from noise? 
Surprisingly, we demonstrate that under some circumstances, quantum noise can be used 
to improve the performance of quantum reservoir computing, a prominent and recent 
quantum machine learning algorithm.  
Our results show that the amplitude damping noise can be beneficial to machine learning, 
while the depolarizing and phase damping noises should be prioritized for correction. 
This critical result sheds new light into the physical mechanisms underlying 
quantum devices, 
providing solid practical prescriptions for a successful implementation of 
quantum information processing in nowadays hardware.
\end{abstract}

\maketitle

\section{Introduction}
Machine learning (ML) is among the most disruptive technological developments of 
the early 21st century \cite{PNAS1,PNAS2}.
However, despite existing ML solutions capable of coping with systems of moderate size, 
learning more complex patterns often requires the use of a large number of parameters 
and long training times; 
this fact conditions its success to having access to high performance computational resources. 
For this reason, a tremendous interest has recently arisen for a technological field 
with potential to dramatically improve many of these algorithms: quantum ML (QML). 
To unravel the full potential of QML algorithms, fault-tolerant computers with millions 
of qubits and low error-rates are needed. Although the actual realization of these 
devices is still decades ahead, the so-called noisy intermediate-scale quantum (NISQ) 
era has been reached. Thanks to NISQ designs, Google recently claimed \cite{Google19} to have achieved quantum supremacy \cite{QSupremacy}, not without controversy \cite{IBM19, no-supremacy}, with the quantum computers available today. Moreover, an experimental demonstration of quantum speed-up on a NP-hard problem regarding Gaussian Boson sampling \cite{BosonSampling} was recently provided.

One of the biggest challenges of the current quantum devices is the presence of noise. 
They perform noisy quantum operations with limited coherence time, 
which affects the performance of quantum algorithms. 
To overcome this limitation, great effort has been devoted to designing error-correcting methods \cite{ErrorCorrecting,ErrorCorrecting2}, which correct the errors in the quantum 
hardware as the algorithm goes on, and also error-mitigation techniques 
\cite{errorMitigation, errorMitigation2}, which aim to reduce the noise of the outputs 
after the algorithm has been executed.  Recently, Google Quantum AI proved the scalability of error-correcting techniques \cite{google2023suppressing}, which is the first step towards fault-tolerant computation.  
Even though these methods can sometimes successfully reduce quantum noise, 
a fundamental question still remains open:
Can the presence of noise in quantum devices be beneficial for quantum machine 
learning algorithms? 

The aim of this paper is to address this issue in a highly relevant NISQ algorithm: 
quantum reservoir computing (QRC) \cite{reviewQRC}.  
This algorithm uses random quantum circuits, carefully chosen from a certain family, 
in order to extract relevant properties from the input data. 
The measurements of the quantum circuits are then fed to a ML model, 
which provides the final prediction. 
This simple learning structure makes of QRC a suitable QML algorithm for NISQ devices. QRs have been used in a wide range of applications, the most common being classical time-series forecasting \cite{time_series1,time_series2,time_series3,time_Series4,time_series5, DynamicalIsing}. Regarding quantum tasks, the method presented in Ref.~\cite{QRC2} involves the detection of entanglement and computation of associated quantities, which are challenging to measure accurately in experimental setups. Additionally, in Ref.~\cite{QuantumTomography}, quantum state tomography using a quantum reservoir has been developed. This method enables reconstruction of the unknown density matrix of the input quantum state with only a single measurement on local observables of the reservoir nodes, without requiring correlation detection. QRs have also been used to compute the preparation of desired quantum states, such as anti-bunched and cat states in Ref.~\cite{StatePreparation1} or maximally entangled states, NOON, W, cluster, and discorded states in Ref.~\cite{StatePreparation2}.

The design of the QR has recentlty proven to be crucial to guarantee optimal 
performance in the ML task \cite{DynamicalIsing,QRCNISQ}. 
However, these studies use \textit{noiseless} quantum simulations, 
which do not take into account the real limitations of current quantum hardware. 
Thus, whether real, noisy implementations of QRs provide advantage over 
classical ML methods is still an open question. 
In Ref.~\cite{noise_VQA} the presence of noise has recently been used 
to improve the convergence of variational quantum algorithms.

In this work, QRs are used to solve a quantum chemistry problem, consisting of predicting the first excited energy of the LiH molecule from its gorund state (see Sect. \ref{sect:QML_task}),
which has become a common benchmark for QML  \cite{Kais, quantumchemQRC, QRCNISQ}. 
Quantum chemistry is one of the areas where quantum computing has highest potential 
of outperforming traditional methods \cite{PNAS2}, since the complexity of the problem 
increases exponentially with the system's degrees of freedom. 
The exponential size of the Hilbert space allows to study high-dimensional systems 
with few quantum computational resources, when compared to classical methods. 
Our results show that certain types of noise can actually provide a better 
performance for QRC than noiseless reservoirs. 
Our numerical experiments are further supported with a theoretical 
demonstration.
Moreover, we provide a practical criterion to decide how to use quantum noise to improve 
the performance of the algorithm, and also what noise should be a priority to correct.

%
\section{Results}
The QML task considered in this work consists on predicting 
the excited electronic energy $E_1$ from the corresponding ground 
state $\ket{\psi_0}_{R}$ with energy $E_0$ for the LiH molecule, using noisy QRs. 
Three noise models are considered in this study: the \textit{depolarizing channel}, 
the \textit{amplitude damping channel} and the \textit{phase damping channel}. 
Full description of the QML task and noise models is provided in 
section \textbf{Methods} below.  
%
\begin{figure}
 \includegraphics[width=0.99\columnwidth]{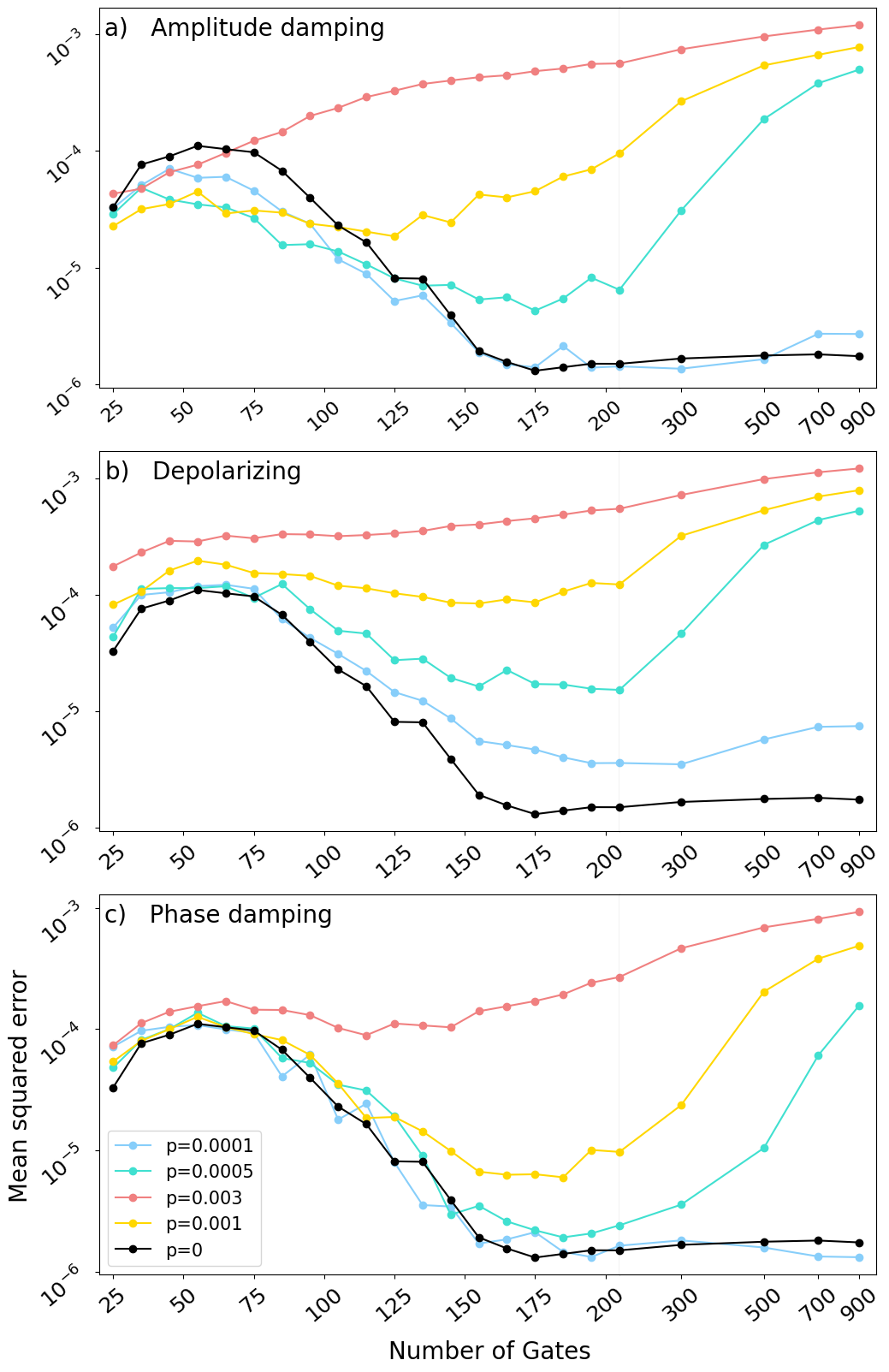}
  \caption{(Averaged) Mean squared error of the quantum reservoirs with 
  a) amplitude damping noise, b) depolarizing noise and c) phase damping noise, 
  as a function of the number of gates of the circuit. 
  Averages are made over 100 simulations. }
 \label{fig:1}
\end{figure}
Figure~\ref{fig:1} shows the mean squared error (MSE) 
in $E_1$ predicted with our QRs as a function of the number of gates, 
for different values of the error probability $p$ (colored curves) 
and noise models (panels),
together with the results for the corresponding noiseless reservoir (in black).

As expected, the general tendency of the MSEs is to grow with the
noise characterized by $p$. However, a careful
comparison of the three plots in Fig.~\ref{fig:1} surprisingly demonstrates that the 
amplitude damping noise renders results which are significantly different from those 
obtained in the other two cases. 
Indeed, if the number of gates and error probability are small enough, 
the QRs with amplitude damping noise provides better results than the noiseless QR.
The same conclusion applies for the higher values of $p$, although in those
cases the threshold number of gates for better performance decreases.
This is a very significant result, since it means that, 
contrary to the commonly accepted belief, the presence of noise is here 
\textit{beneficial} for the performance of the quantum algorithm,
and, more importantly, it takes place within the limitations of the NISQ era.
As an example, for $p=0.0005$ (green curve) all noisy reservoirs render
better performance than the noiseless counterpart when the number of gates 
is smaller than $135$. Current quantum processors typically have error rates around $p=0.001$, which are expected to be significantly reduced soon by employing error-correction techniques \cite{google2023suppressing}. 

A practical criterion to decide when noise can be used to improve the 
performance of QRC is provided in Table~\ref{tab:optimal_fidelity},
which shows the averaged fidelity between the output noisy state $\rho$ and the 
noiseless state $\ket{\psi}$ for the circuits subjected to an amplitude damping noise 
with different values of the error probability. 
The number of gates has been chosen to be as large as possible provided that 
the noisy reservoirs outperform the noiseless ones. 
These results imply that when the fidelity is greater than $0.96$, 
the noisy reservoirs outperform the noiseless ones at the QML task, and accordingly the noise should \textit{not} be corrected. 
Finally, also notice that for $p=0.0001$ the fidelity is always higher than $0.96$, 
and thus the performance of the noisy QRs is always higher or equal than their noiseless counterparts. Table \ref{tab:optimal_fidelity} also shows that the number of gates needed to outperform the noiseless reservoirs is of the order of 100 quantum gates, which corresponds to an average circuit depth of 10-15 gates. Recently, there have been multiple applications of quantum machine learning algorithms using shallow quantum circuits of similar depth. In particular, multiple shallow quantum neural networks have been combined to solve six benchmark classification problems in Ref. \cite{Qnet}. Also, a hybrid quantum-classical graph neural network, which used quantum circuits of depth$<$10, was developed for particle track reconstruction in particle acceleration experiments \cite{hybridGNN}. Finally, in Ref. \cite{quantumDatasets}, the authors propose the design of quantum datasets for quantum machine learning tasks, where the classification label is encoded in the amount of entanglement of the quantum states. Their results shows that the quantum datasets are successfully implemented with circuit depth smaller than 7. Therefore, quantum circuits with depths of 10-15 gates can provide useful applications in various domains. This is the regime where the amplitude damping noise provides an advantage over noiseless quantum circuits in our setting, which suggests that this type of noise may be beneficial to other quantum machine learning tasks. The extension of this analysis to other applications, such as time series forecasting, will be explored in future works.  
%
\begin{table}
    \centering
    \begin{tabular}{ccc}
        \hline \hline \\[-2.5ex]
        Error prob. & Optimal    & Fidelity \\  [-0.5ex]    
               $p$  & \# of gates   &  (averaged)\\ [1ex]
        \hline
        0.0001      &  150          &  0.990\\
        0.0005      &  135          &  0.965 \\
        0.0010      &  105           &  0.956\\
        0.0030      &  65           &   0.962\\
        \hline
    \end{tabular}
    \caption{(Averaged) Fidelity between the noisy and noiseless final quantum states 
    for the circuits with amplitude damping noise (see text for details). 
    The number of quantum gates is chosen so that the performance of the noisy 
    reservoirs outperforms that of the noiseless reservoirs.}
    \label{tab:optimal_fidelity}
\end{table}

A second conclusion from the comparison among plots in Fig.~\ref{fig:1} is that
the behavior for depolarizing and the phase damping channels is significantly 
different than for the amplitude damping one. 
In the former cases, the performance of the noisy reservoirs is always worse than 
that of the noiseless one, even for small error probabilities. 

A third result that can be extracted from our calculations is that the 
tendency of the algorithm performance when the reservoirs have a large 
number of gates is the same for the three noise models considered 
(except for the smallest value of $p=0.0001$). 
While the performance of the noiseless reservoirs stabilizes to a constant value 
as the number of gates increases, the noisy reservoirs decrease their performance,
seemingly going to the same growing behavior.  
This is due to the fact that the quantum channels are applied after each gate, 
and thus circuits with a large number of gates have larger noise rates, 
which highly decreases the fidelity of the output state. 
For this reason, even though increasing the number of gates has no effect in 
the noiseless simulations, it highly affects the performance of the noisy circuits, 
and thus the number of gates should be optimized in this case. 

Having analyzed the MSE results, we next provide a theoretical 
explanation for the different behavior of the three noisy reservoirs. 
In the first place, the depolarizing and phase damping channels give similar results, 
except that the performance of the former decreases faster than that for the latter. 
This effect can be explained with the aid of Table~\ref{tab:average_fidelity}, 
where the averaged fidelity of each error model over the first 200 gates
is given. 
\begin{table}
    \centering
    \begin{tabular}{cccc}
        \hline \hline \\ [-2.5ex]
        Error prob. &  Amplitude  &  Depolarizing  &  Phase  \\ [-0.5ex]
            $p$     &    damping  &                & damping \\ [1ex] 
        \hline
        0.0001      & 0.995       & 0.994          &  0.998  \\
        0.0005      & 0.975       & 0.971          &  0.988  \\
        0.0010      & 0.951       & 0.944          &  0.976  \\
        0.0030      & 0.862       & 0.842          &  0.931  \\
        \hline
    \end{tabular}
    \caption{(Averaged) Fidelity between the noisy and noiseless final quantum states 
    for the circuits with the three noise models. 
    Fidelity is averaged over all the quantum reservoirs with less than 200 gates, 
    with the same noise model.}
    \label{tab:average_fidelity}
\end{table}
As can be seen, the depolarizing channel decreases the fidelity of the output 
much faster than the phase damping, which explains the different tendency 
in the corresponding ML performances. 
On the other hand, the amplitude damping channel is the only one that can improve the
performance of the noiseless reservoirs in the case of few gates and small error rates. 
The main difference between amplitude damping and the other channels is that the former 
is not unital, i.e.~it does not preserve the identity operator. 

Let us consider now how this fact affects the distribution of noisy states 
in the Pauli space. 
For this purpose, let $\rho'$ be the $n-qubit$ density matrix obtained after applying $N-1$ noisy gates, 
(with the noise described by the quantum channel $\epsilon$), 
and then apply the $N$-th noisy gate $U$. 
The state becomes $\epsilon(\rho)$, defined as:
\begin{equation}
  \epsilon(\rho) = \sum_{m=1} M_m \rho M_m^\dagger, \quad \rho = U \, \rho' \, U^\dag,
\end{equation}
where $\rho$ is the state after applying gate $U$ \emph{without} noise.
Now, both $\rho$ and $\epsilon(\rho)$ can be written as linear combinations of Pauli basis
operators $\{P_i\}_i$, where each one of them is the tensor product of the
Pauli operators $\{ X,Y,Z,\mathbb{I}\}$ as
\begin{eqnarray}
   &&  \rho = \sum_i a_i P_i, \quad \text{with }a_i = \frac{1}{2^n} \tr (P_i \rho), \\
   && \epsilon(\rho) =\sum_i b_i P_i, \quad \text{with }b_i = \frac{1}{2^n}\tr [P_i \epsilon(\rho)].
\end{eqnarray}
Notice here that some of the coefficients $b_i$ will be used to feed the ML model after 
applying all the gates of the circuit and make the final predictions. 
Thus, expanding the final quantum states in this basis is suitable to understand 
the behavior of the QRC algorithm. 
Next, we study the relation between coefficients $\{a_i\}$ and $\{b_i\}$. 
Since the operators $P_i$ are tensor product of Pauli operators, 
it is sufficient to study how each of the noise models $\epsilon$ maps the four Pauli operators. 
The results are shown in Table~\ref{tab:quantum_channel}, 
where we see that $\epsilon(P_i)$ is always proportional to $P_i$, 
except for $\epsilon(\mathbb{I})$ with the amplitude damping channel. 
Indeed, it is for this reason that, with depolarizing or phase damping noises,
the quantum channel only mitigates coefficients in the Pauli space. 
On the other hand, the amplitude damping channel can introduce additional non-zero 
terms to the Pauli decomposition. 
Also, this explains why, for low noise rates, the shapes of the MSE curves 
for depolarizing and phase damping are similar to that for the noiseless scenario, 
but not for the amplitude damping one. 
Table~\ref{tab:quantum_channel} also explains why the phase damping channel 
provides states with higher fidelity than the depolarizing channel. 
The phase damping channel leaves the $Z$ operator invariant, 
and also produces lower mitigation of the $X$ and $Y$ coefficients compared 
to the depolarizing channel. 
For this reason, even though both the depolarizing and phase damping channels are unital, 
the depolarizing channel decreases the ML performance faster, 
and its correction should be prioritized.
 \begin{table}
     \centering
     \begin{tabular}{cccc}
     \hline \hline                                     \\[-2.5ex]
          &  Amplitude & Depolarizing  & Phase         \\ [-0.5ex]
          & damping    &               & damping       \\ [1ex]
          \hline 
       $\epsilon(X)$   & $\sqrt{1-p}\;X$ & $(1-\frac{4}{3}p)X$ & $(1-p)\;X$ \\
       $\epsilon(Y)$   & $\sqrt{1-p}\;Y$ & $(1-\frac{4}{3}p)Y$ & $(1-p)\;Y$ \\
       $\epsilon(Z)$   & $(1-p)\;Z$      & $(1-\frac{4}{3}p)Z$ & $Z$ \\
       $\epsilon(\mathbb{I})$   & $\mathbb{I} + pZ$ & $\mathbb{I}$ & $\mathbb{I}$ \\
       \hline
     \end{tabular}
     \caption{Expressions for the error channel $\epsilon$ when applied 
     to the four basis Pauli operators.}
     \label{tab:quantum_channel}
 \end{table}

Let us provide a mathematical demonstration for this fact. 
For any Pauli operator $P_i$, the coefficient in the Pauli space with the depolarizing 
and phase damping channels is
 \begin{equation}
     b_i = \frac{1}{2^n}\tr[P_i\;\epsilon(\rho) ] = \frac{1}{2^n}\alpha_i \;tr(P_i  \rho ) = \alpha_i \; a_i, 
        \quad 0 \leq \alpha_i \leq 1,
 \end{equation}
 and therefore the noisy channel mitigates coefficient $a_i$. 
 However, let us take a gate with amplitude damping noise. 
 Suppose channel $\epsilon$ acts non-trivially on qubit $j$, that is, 
 the Kraus operators for $\epsilon$  are of the form 
 $\tilde{M}_m = \mathbb{I} \otimes \cdots \otimes M_m \otimes \mathbb{I} \otimes \cdots \mathbb{I}$, 
 with $M_m$ in the $j$-th position. 
 Suppose now that we measure $P_i$ (the $i$-th operator in the Pauli basis  associated  
 to coefficient $a_i$),  where $P_i$ acts as a $Z$ operator on the $j$-th qubit 
 ($P_i = P^0 \otimes \cdots P^{j-1} \otimes Z \otimes P^{j+1} \otimes \cdots P^n$). Let's also take $P_k=P^0 \otimes \cdots \otimes P^{j-1} \otimes \mathbb{I} \otimes P^{j+1} \otimes \cdots \otimes P^n$, with $a_k$ associated to $P_k$. 
 Then, the coefficient $b_i$ is
 \begin{equation}
    \begin{array}{rl}
     b_i  = &\displaystyle\frac{1}{2^n} \tr[P_i\epsilon(\rho)] = 
     \frac{1}{2^n} \sum_l a_l \tr[P_i \epsilon(P_l)]  \nonumber \\
    =& \displaystyle\frac{1}{2^n} \Big(a_i \tr[P_i \epsilon(P_i)] 
                   +  a_k \tr[P_i \epsilon(P_k)]\Big) \nonumber \\
    =&\displaystyle\frac{1}{2^n}\Big(a_i (1-p) \tr[P_i^2] + a_k\tr[P_i(P_k + pP_i)]\Big)\nonumber \\
    =& (1-p)a_i + p a_k
    \end{array} 
    \end{equation}
When $a_i=0$ but $a_k \neq 0$, the coefficient $b_i$ is different from 0, 
and thus the amplitude damping noise introduces an extra coefficient in the Pauli space.
Therefore, we can conclude that the amplitude damping channel allows to introduce 
additional non-zero coefficients in the Pauli space, instead of only mitigating them.
For this reason, for $p$ small enough, the amplitude channel can introduce new 
non-zero terms in the Pauli space without mitigating too much the rest of them. 
    
\begin{figure}
  \includegraphics[width=0.95\columnwidth]{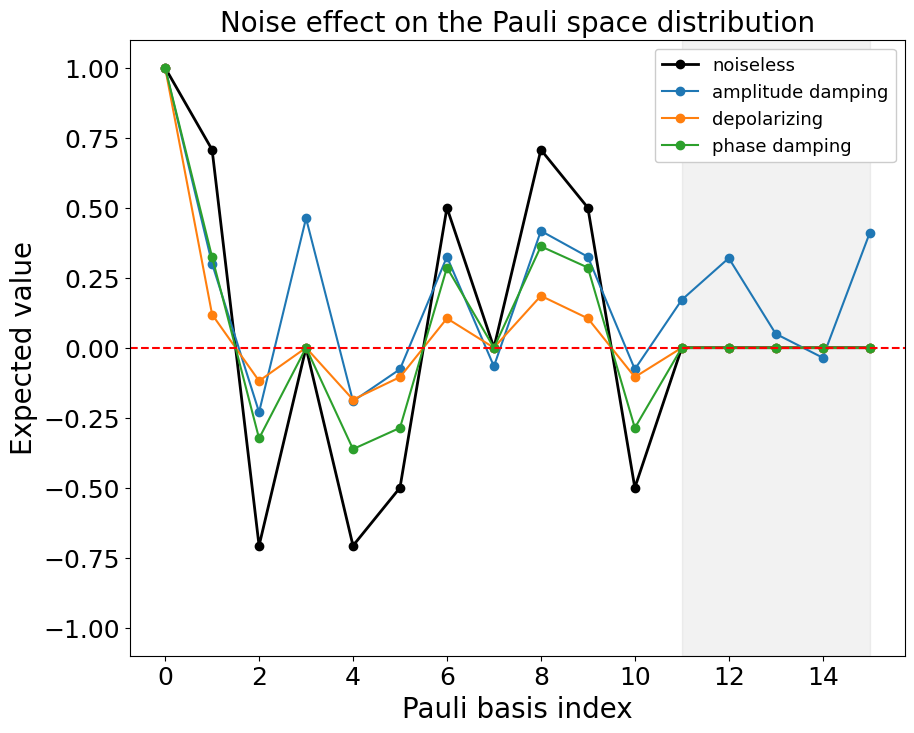}
  \caption{Coefficients in the Pauli space of a 2-qubits toy model
  (see text for motivation) consisting of a random quantum circuit 
  with 10 gates from the G3= \{H, CNOT, T\} family and error probability $p=0.2$, 
  for the three noise models studied in this work together 
  with the noiseless coefficients in black.} 
\label{fig:2}
\end{figure}

\begin{figure}[b]
  \includegraphics[width=0.85\columnwidth]{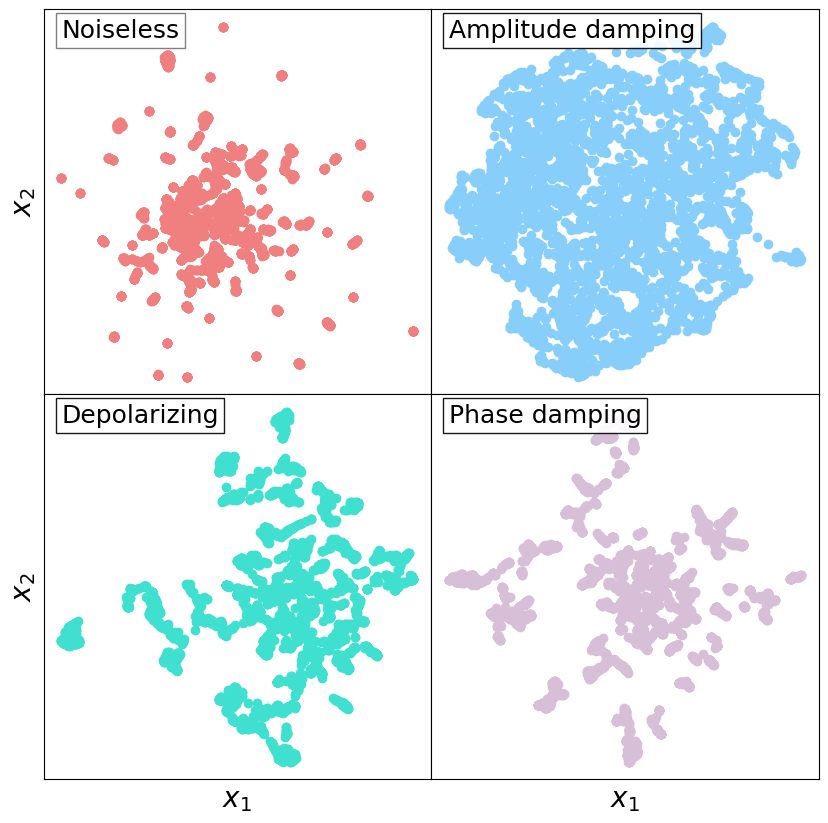}
  \caption{Reduced 2D (from 16D) representation of the distribution in
  the Pauli space of 400 simulations of the toy model of Fig.~\ref{fig:2}. 
  Variables $x_1$ and $x_2$ are selected using the UMAP algorithm 
  of Ref. \cite{UMAP2}.}
\label{fig:3}
\end{figure}

The previous theorem can be further illustrated with 
a two qubits toy model example.
  We design a QR with the three different quantum noise models and calculate the 
  distribution of the Pauli coefficients at the end of the circuit. 
  Figure~\ref{fig:2} shows the outcomes of the measurements 
  for a random circuit with 10 gates and an error rate of $p=0.2$. 
  We see that all noise models mitigate the non-zero coefficients. 
  However, the shadowed area shows a region where the noiseless simulation
  (as well as the depolarizing and phase damping simulations) give
  zero expectation values. 
  More importantly, the amplitude damping circuit has non-zero expectation  
  values for the same operators, which means that this quantum channel has introduced 
  non-zero terms in the Pauli distribution.
  For small error rates, the noisy quantum reservoirs provide better performance, 
  since having amplitude damping noise produces a similar effect in terms of performance of the QRs as having more 
  quantum gates in the circuit, as can be seen in Fig. \ref{fig:1} and also in Fig. 3 from Ref. \cite{QRCNISQ}. 
  To better visualize this effect, we design 4000 random circuits and see how the 
  final state $\rho$ fills the Pauli space. 
  Since the Pauli space in the 2-qubit system is a 16-dimensional space, 
  we use a dimensionality reduction technique called UMAP \cite{UMAP2} to visualize 
  the distribution in 2D.  
  The results are shown in Fig.~\ref{fig:3}. 
  We see that the amplitude damping channel fills the Pauli space faster 
  than the other circuits, including the noiseless QR, 
  thus confirming the hypothesis that the amplitude damping channel acts equivalently 
  as having more quantum gates. 


\section{Conclusions}
In this paper, 
the effect on the QRC performance of three different paradigmatic noise models,
effectively covering the most relevant and widely studied channels \cite{nielsen2010quantum,majorizationBenchmark} affecting quantum devices, are evaluated.
Contrary to common belief, we demonstrate that, under certain circumstances, noise, which constitutes the biggest challenge for quantum computing and QML, can be beneficial for quantum reservoir computing.
Remarkably, we show that for error rates $p \lesssim 0.0005$ 
or state fidelities of at least $0.96$, the presence of an amplitude damping channel renders better performance than noiseless QRs for the QML task at study, which consists of predicting the excited energy of the LiH molecule from its ground state. Moreover, our theoretical demonstration suggests that the benefits of amplitude damping noise will be also present in other tasks involving QRs, such as time series forecasting, which will be explored in future works. The performance of error mitigation techniques applied to QRs with different noise models will also be investigated in future studied. 
The effect of the amplitude damping channel is explained by analyzing the distribution in the Pauli space of the resulting density matrices after suffering the amplitude damping noise. This channel introduces additional non-zero coefficients in the Pauli space, 
which produces a similar effect as having more quantum gates in the original circuits. On the other hand, the depolarizing and phase damping channels only reduce the amplitude of the coefficients in the Pauli space, this producing poorer results. 
The depolarizing channel is the one that mitigates fastest these values, so our prescription is that its correction should be a priority. For this reason, error-correcting methods that target depolarizing noise \cite{DeepQRL} should be employed in machine learning tasks involving the use of QRs. 

\section{Methods}

subsection{Quantum Reservoir Computing}
The idea of QRC lies in using a Hilbert space as an enhanced feature space of the input data. In this way, the feature space enhanced by quantum entangling operations are used to feed a classical machine learning model, which predicts the desired target. Consider a dataset $\{(x_i,y_i)\}$, where $x_i$ are the input samples and $y_i$ the target outputs. The data samples are encoded as an $n$-qubit quantum state $\ket{x}_i$. Then, a \emph{random} unitary transformation $U$ is applied to extract features from the input data, resulting in the quantum state $U \ket{x}_i$. The operator $U$ is sampled from a carefully selected family of operators, such that $U$ creates enough entanglement to generate useful transformations of the input data while being experimentally feasible. For this reason, the design of the reservoir $U$ is crucial for the optimal performance of the algorithm. In this work, the QRs are designed as random quantum circuits whose gates are chosen from a finite set. In a previous work \cite{QRCNISQ}, it was proven that the majorization principle \cite{majorization_original}, a relevant indicator of quantum circuit complexity \cite{majorization} for the NISQ era, serves as an indicator of performance for QRs. The G3=\{CNOT,H,T\} family of random quantum circuits was found to provide an optimal design for QRs, where CNOT is the controlled-NOT gate, H stands for Hadamard, and T is the $\pi/8$ phase gate. For this reason, the G3 family is used to generate the reservoirs $U$ with a fixed number of gates.\\
\\
After applying $U$ to the initial quantum state, the expected value of single-qubit Pauli observables is measured, providing the extracted features $X$. Such features are fed to a classical machine learning algorithm. Even though complex machine learning models can be used, the QR should be able to extract valuable features so that a simple machine learning model can predict the targets $y_i$. For this reason, a linear model is usually used to learn the output.

\subsection{Quantum machine learning task}
\label{sect:QML_task}
In this work, QRs are used to predict the first excited electronic energy $E_1$ 
using only the associated ground state $\ket{\psi_0}_{R}$ with energy $E_0$ for the LiH molecule. 
The ground state $\ket{\psi_0}_{R}$ for the LiH Hamiltonian is calculated by exact 
diagonalization for different values of the internuclear distance $R \in [0.5, 3.5]$ a.u. The details of the ground state calculation are given in Ref~\cite{QRCNISQ}. 
For this case, $n=8$ qubits are needed to describe the ground state, and QRs are used to predict the relative excited energy $\Delta E(R)$. The dataset $\{\ket{\psi_0}_{R}, \Delta E(R)\}_R$ is split into training and test sets, where the test set contains the 30\% of the data $R \in [1.1, 2.0]$ a.u., and it is designed so that the QML algorithm has to extrapolate to \textit{new} data samples. 

\subsection{Noise models}
The goal of this work is to study the effect of three noise models on the performance of the 
ML task, for different error probabilities and number of quantum gates. 
It is important to note that these models embody the overwhelming majority of noise types to which modern hardware is subjected to, this pointing out to the generality of our
conclusions. The first noise model that we consider is the \textit{amplitude damping channel}, which reproduces the effect 
of energy dissipation, that is, the loss of energy of a quantum state to its environment. 
It provides a model of the decay of an excited two-level atom due to the spontaneous emission 
of a photon with probability $p$. 
The Kraus operators of this channel are given by
\begin{equation}
    M_0 = 
    \begin{pmatrix}
    1 & 0 \\
    0 & \sqrt{1-p}
    \end{pmatrix}, \quad
    M_1 = 
    \begin{pmatrix}
    0 & \sqrt{p} \\
    0 & 0
    \end{pmatrix}.
\end{equation}
 The operator $M_1$ transforms $\ket{1}$ to $\ket{0}$, which corresponds to the process 
 of losing energy to the environment. 
 The operator $M_0$ leaves $\ket{0}$ unchanged, but reduces the amplitude of $\ket{1}$. 
 The quantum channel is thus
 \begin{equation}
     \epsilon(\rho) = M_0 \, \rho \, M_0^\dag + M_1 \, \rho \, M_1^\dag 
     = \begin{pmatrix}
     \rho_{00} + p \; \rho_{11} & \sqrt{1-p} \; \rho_{01}\\
     \sqrt{1-p}\; \rho_{10} & (1-p) \; \rho_{11}
     \end{pmatrix} .
 \end{equation}
 The second noise model is described by the \textit{phase damping channel}, 
 which models the loss of quantum information without loss of energy. 
 The Kraus operators for the process are
 \begin{equation}
     M_0 = \sqrt{1-p} \;\; \mathbb{I}, \quad M_1 =  \begin{pmatrix}
     \sqrt{p} & 0\\
     0 & 0
     \end{pmatrix}, 
     \quad M_2 = \begin{pmatrix}
     0 & 0 \\
     0 & \sqrt{p}
     \end{pmatrix} ,
 \end{equation}
 and the quantum channel is then
 \begin{eqnarray}     
    \epsilon(\rho) &=& M_0 \, \rho \, M_0^\dag + M_1 \, \rho \, M_1^\dag 
                        + M_2 \, \rho \, M_2^\dag 
     \nonumber \\
     &=& \left(1-\frac{p}{2}\right) \, \rho + \frac{p}{2} \; Z\, \rho \, Z.
 \end{eqnarray}
An alternative interpretation of the phase damping channel is that the state 
 $\rho$ is left intact with probability $1-p/2$, and a $Z$ operator is applied 
 with probability $p/2$. 
 The last noise model is described by the \textit{depolarizing channel}. 
 In this case, a Pauli error $X$, $Y$ or $Z$ occurs with the same probability $p$. 
 The Kraus operators are
 \begin{equation}
     M_0 = \sqrt{1-p}\,\mathbb{I}, \;M_1 = \sqrt{\frac{p}{3}}  X, \; 
     M_2 = \sqrt{\frac{p}{3}} Y, 
     M_3 = \sqrt{\frac{p}{3}} Z,
 \end{equation}
 and the quantum channel is
 \begin{equation}
     \epsilon(\rho) = (1-p)\, \rho + \frac{p}{3} \, (X\rho X + Y \rho Y + Z \rho Z) 
     = (1-p) \, \rho + \frac{p}{2} \; \mathbb{I}.
 \end{equation}
  The depolarizing channel transforms the state $\rho$ into the maximally mixed state 
  with probability $p$. 
  Notice that the amplitude damping channel is the only one which is not \textit{unital}, 
  since it does not map the identity operator into itself. 
  In general terms, it belongs to the kind of volume contracting environments in
  phase space with many generalizations that include the quantization of classical friction. 

  \subsection{Training process}
  
  The training steps of the algorithm are the following. 
  First, the quantum circuit is initialized with the molecular ground state $\ket{\psi_0}_{R}$ 
  for a certain configuration $R$. 
  Next, a noisy quantum circuit with fixed number of gates is applied to $\ket{\psi_0}_{R}$. 
  Then, we measure the local Pauli operators $\{X_0,  Z_0, \cdots, X_n,  Z_n\}$, 
  where $X_i,Z_i$ are the Pauli operators $X,Z$ applied to the $i$-th qubit, 
  thus obtaining the vector
\begin{equation}
    X(R) = \left( \expval{X_0},  \expval{Z_0}, \cdots,  
        \expval{X_n}, \expval{Z_n} \right)^T 
\end{equation}
which provides the extracted information from the ground state. 
Recall that for a noisy state $\rho$, the expectation value of an operator $P$ 
is given by 
$\expval{P} = \tr(P \rho)$. 
The vector $X(R)$ is fed to a classical machine learning algorithm, 
in this case a ridge regression, which is a linear model with $L^2$ regularization. 
The optimal regularization parameter was $\alpha = 10^{-9}$, 
which reduces overfitting while maintaining optimal prediction capacity \cite{QRCNISQ}. 
The effect of the different noise channels in the algorithm performance is studied 
by varying the error probability $p$. 
We perform 100 simulations for probabilities $p=0.0001, 0.0005, 0.001, 0.003$ 
for each quantum channel, and compare the performance of the model with 
the noiseless simulation ($p=0$). 
We also study how the number of quantum gates affects the performance of the reservoirs. 
We design circuits varying the number of gates from 25 to 215 in intervals of 10 gates. 
Also, we study the performance for large number of quantum gates, 
using 300, 500, 700 and 900 of them. All the simulations have been performed using Qiskit software \cite{Qiskit} via exact quantum circuit simulation with custom noise models (see Sect. \ref{sect:code}). 

\section{Data availability}
The datasets generated and analysed during the current study are available in the 
GitHub repository, \href{https://github.com/laiadc/Optimal\_QRC}
{https://github.com/laiadc/Optimal\_QRC\_noise}.

\section{Code availability}
\label{sect:code}
The underlying code for this study is available in the Github repository and can be accessed 
via this link \href{https://github.com/laiadc/Optimal\_QRC\_noise}
{https://github.com/laiadc/Optimal\_QRC\_noise}.

\section{Competing interests}
The authors declare no competing financial or non-financial interests.

\section{Author contributions}
All authors developed the idea and the theory. LD performed the calculations and analyzed the data. 
All authors contributed to the discussions and interpretations of the results and wrote the manuscript.
\section{Acknowledgments}
The project that gave rise to these results received the support of a fellowship from 
``la Caixa'' Foundation (ID 100010434). The fellowship code is LCF/BQ/DR20/11790028.
This work has also been partially supported by the 
Spanish Ministry of Science, Innovation and Universities, 
Gobierno de Espa\~na, under Contracts No.\ PID2021-122711NB-C21, ICMAT Severo Ochoa CEX2019-000904-S.  
The funders played no role in study design, data collection, analysis and interpretation of data, 
or the writing of this manuscript. 

\bibliography{bibliography}

\end{document}